\pdfoutput=1
\documentclass[twocolumn,twoside]{article}
\usepackage[backend=biber, natbib=true, style=authoryear-comp, sorting=none, maxnames=3, maxalphanames=3]{biblatex}
\usepackage[a4paper,includeheadfoot,margin=2cm,top=1.5cm,bottom=2cm]{geometry}
\usepackage{amsmath}
\usepackage{amssymb}
\usepackage{amsthm}
\usepackage{graphicx}
\usepackage{float}
\usepackage{dblfloatfix}
\usepackage{fixltx2e}
\usepackage{adjustbox}
\usepackage{subcaption}
\usepackage{hyperref}
\usepackage{gensymb}
\usepackage[utf8]{inputenc}
\usepackage{tikz}
\usetikzlibrary{tikzmark}
\usepackage{authblk}
\usepackage[font=small,labelfont=bf]{caption}

\usepackage{mathtools}

\DeclareMathOperator*{\argmin}{\arg\!\min}
\DeclareMathOperator*{\kernel}{Ker}
\DeclareMathOperator*{\image}{Im}

\newcommand{\Mod}[1]{\ (\mathrm{mod}\ #1)}

\newtheorem{thm}{Theorem}
\newtheorem{defn}{Definition}

\newcommand{\A}{\mathbb{A}}
\newcommand{\R}{\mathbb{R}}
\newcommand{\Z}{\mathbb{Z}}
\newcommand{\F}{\mathbb{F}}
\newcommand{\I}{\mathbb{I}}

\newcommand{\N}{\mathcal{N}}
\newcommand{\X}{\mathcal{X}}

\graphicspath{{./plots/}}

\title{Decoding of neural data using cohomological feature extraction}

\author[a]{Erik Rybakken}
\author[a]{Nils Baas}
\author[b]{Benjamin Dunn}

\affil[a]{\{erik.rybakken, nils.baas\}@ntnu.no\\ Department of Mathematical Sciences, Norwegian University of Science and Technology, 7491 Trondheim, Norway}
\affil[b]{benjamin.dunn@ntnu.no\\ Kavli Institute for Systems Neuroscience, Norwegian University of Science and Technology, 7491 Trondheim, Norway}

\begin{document}

\twocolumn[
  \begin{@twocolumnfalse}
    \maketitle
    \begin{abstract}
We introduce a novel data-driven approach to discover and decode features in the neural code coming from large population neural recordings with minimal assumptions, using cohomological feature extraction. 
We apply our approach to neural recordings of mice moving freely in a box, where we find a circular feature. 
We then observe that the decoded value corresponds well to the head direction of the mouse.
Thus we capture head direction cells and decode the head direction from the neural population activity without having to process the behaviour of the mouse. 
Interestingly, the decoded values convey more information about the neural activity than the tracked head direction does, 
with differences that have some spatial organization.
Finally, we note that the residual population activity, after the head direction has been accounted for, retains some low-dimensional structure which is correlated with the speed of the mouse.
    \end{abstract}
\bigskip
  \end{@twocolumnfalse}
]
\section*{Introduction}

\begin{figure*}[t]
\adjustbox{valign=b}{%
  \begin{minipage}[t]{11.4cm}
  \includegraphics[width=\textwidth]{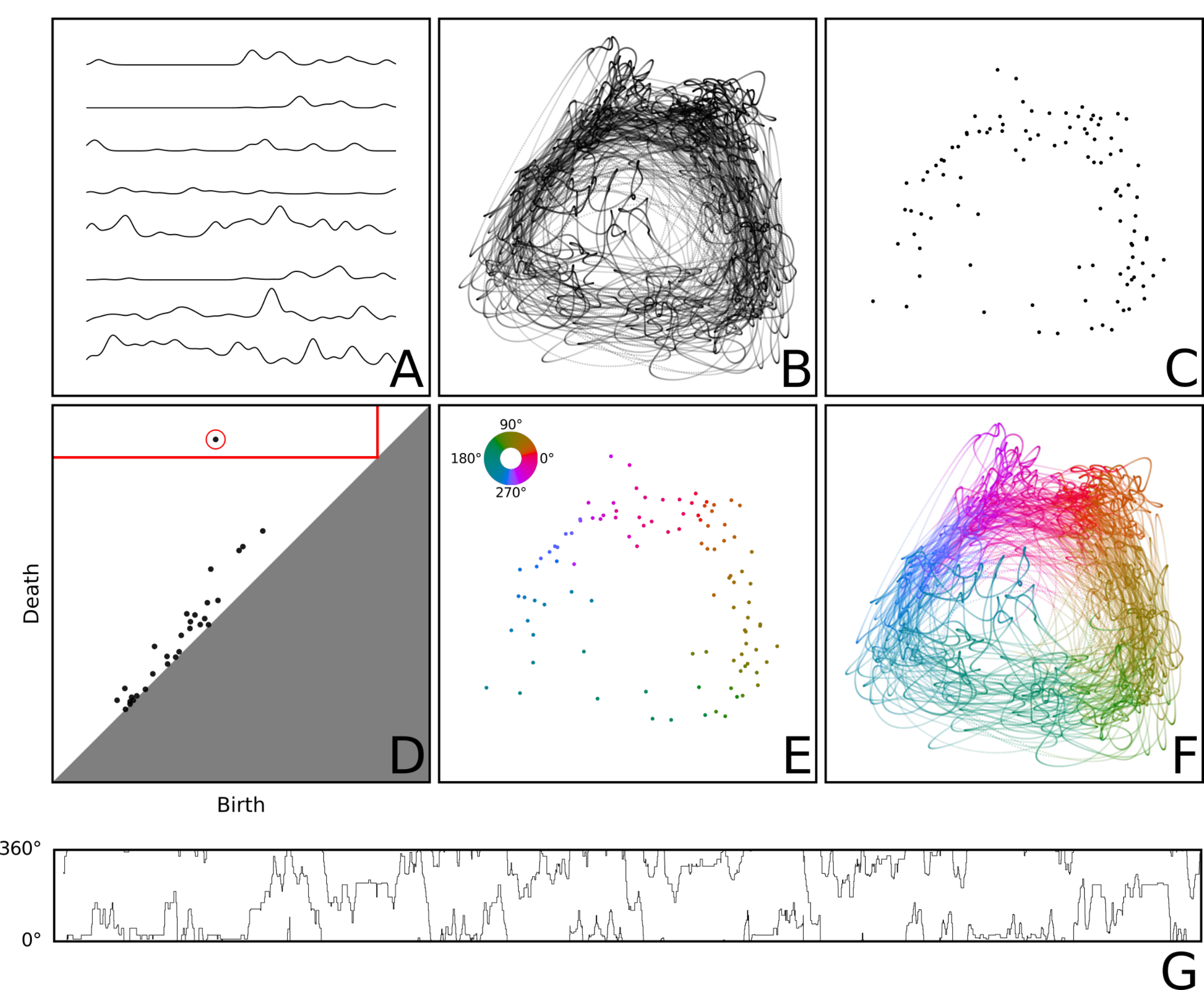}
\end{minipage}}%
\hspace{0.3cm}
  \adjustbox{valign=b}{%
    \begin{minipage}[t]{.34\linewidth}
  \caption{The decoding procedure. We start with an estimated firing rate for each neuron (A) obtained by smoothing the spike trains of the neurons.
    The firing rates are sampled at fixed time intervals, giving a point cloud (B), which is then simplified (see Appendix \ref{simplification}), obtaining a reduced point cloud (C). 
The first persistence diagram (D) of the reduced point cloud is computed. We select the longest living feature (circled in red) and a scale where this feature exists (red lines). 
Circular parametrization is used to obtain a circular map (E) on the reduced point cloud, where the color of a point represents its circular value according to the color wheel in the upper left corner (the same coloring scheme will be used throughout the rest of the article). The map is extended to the full point cloud (F) by giving each point the value of its closest point in the reduced point cloud. 
In (G) we show the decoded circular value as a function of time for the first 20 minutes of the recording, where the circular value at a given time point is the value of the corresponding point in (F). The point clouds in B, C, E and F are displayed as 2D projections.}
\label{method}
\end{minipage}
}
\end{figure*}
  
The neural decoding problem is that of characterizing the relationship between stimuli and the neural response. For example, head direction cells \citep{ranck85} respond with an elevated activity whenever the animal is facing a specific direction. To be able to determine this relationship, however, an experiment must be designed such that the relevant behavior of the animal can be properly sampled and tracked. Accurate tracking of the animal’s complete behavior, including movements of the eyes, whiskers, head, body and limbs is extremely difficult in any freely behaving situation; external cues such as sounds in the frequency range of the animal, odors, visual cues, etc. are also similarly difficult to account for appropriately and yet all of these aspects of the experiment are just a subset of the potential features driving the activity of the neurons. This process of identifying the relevant features is known as model selection (for recent examples see \cite{hardcastle2017multiplexed, mimica2018efficient}). As it is typically performed, model selection indicates the most likely feature or combination of features behind the activity of the neurons. It does not, however, indicate a feature that is not explicitly included. We would therefore prefer to skip the problem of devising and testing all the infinite ways of tracking and processing the behavioral data and instead allow the neural data to speak for themselves, i.e. data-driven model selection.

Data-driven approaches such as ours have been made possible by recent advances in recording technology that permit simultaneous recordings of a large number of neurons. While the possible states of these neurons form a very high-dimensional space, one expects that the neural activity can be described by a smaller set of parameters \citep{tsodyks99}. Standard dimensionality reduction techniques such as principal component analysis (PCA) and factor analysis (see \cite{cunningham2014} for a comprehensive review) can be used to obtain a low-dimensional version of the data. However, most of these methods give us data lying in some euclidean space, while often a different space would be more appropriate. For instance, if the neurons encode head direction, the population activity should be confined to points on a circle, corresponding to specific directions in the room. If the neurons are tuned to more complex features, or perhaps a combination of features, then spaces with richer topology would be appropriate. Since these spaces often have non-trivial topology, we argue that topological methods should be used in addition to traditional methods. Here we use \emph{persistent cohomology} and \emph{circular parametrization} \citep{deSilva2011,johansson2015} to identify the shape of the underlying space and then decode the time-varying position of the neural activity on it.

In summary, our method consists of combining four key steps:

\begin{enumerate}
  \item Dimensional reduction using PCA
  \item Feature identification using persistent cohomology
  \item Decoding using circular parametrization
  \item Removing the contribution of the decoded features on the data using a generalized linear model (GLM)
\end{enumerate}

This procedure is quite general and may qualify to be called \emph{cohomological feature extraction}. Steps 1-3 are illustrated in Figure \ref{method}. In step 4, the process of removing the contribution of a given feature results in a new data set to which the method can then be re-applied, in a way similar to \cite{spreemann15}, in order to characterize additional features in the neural activity.

As an example, we apply this method to neural data recorded from freely behaving mice \citep{peyrache2015, peyrache15data}, discover a prominent circular feature and decode the time-varying position on this circle. We demonstrate that this corresponds well with the head direction, but with subtle, interesting differences.
Finally, we consider the remaining features, following the removal of the head direction component, and find that there is still a structure there which is correlated to the speed of the mouse.

\section*{Method}
\label{methods}

\paragraph{Decoding procedure}

We will here give the details of the decoding procedure summarized in Figure \ref{method}. We are given a set \(\N = \{1,2,\dots,N\}\) of neurons and a spike train \((s^i_1, s^i_2, \dots, s^i_{n^i})\) for each neuron \(i \in \N\), consisting of the time points when the neuron fires\footnote{The spike trains that we analyzed had a duration of about 30-40 minutes.}. The neurons having a mean firing rate lower than 0.05 spikes per second are discarded. The remaining spike trains are smoothed with a gaussian kernel of standard deviation \(\sigma = 1000ms\), and then normalized to take values between 0 and 1. This gives us estimated normalized firing rates \(f_i \colon \R \to [0,1]\) for each neuron \(i \in \N\), as shown in Figure \ref{method}A. The firing rates are then sampled at fixed time intervals of length \(\delta=25.6 ms\)\footnote{Different interval lengths are possible, but we chose the same interval length as were used for the camera tracking for practical purposes.}, resulting in a sequence \((\mathbf{\overline{x}}_0, \mathbf{\overline{x}}_1, \dots )\) of points in \(\R^N\), where \(\mathbf{\overline{x}}_t = (f_1(\delta t), f_2(\delta t), \dots, f_N(\delta t))\). In order to reduce noise, we project this point cloud onto its first \(d=6\) principal components\footnote{Care should be taken here, as is also noted in section 5.1 of \cite{johansson2015}, since if the projecting dimension \(d\) is too small, the projection could give rise to intersections of the underlying space, giving topological artifacts. It lies in our assumptions that the underlying shape is not too complex and is still an embedding when projected to 6 principal components.}, resulting in a sequence \(\X = (\mathbf{x}_0, \mathbf{x}_1, \dots )\) of points in \(\R^d\) as shown in Figure \ref{method}B (in 2D). We then simplify\footnote{See Appendix \ref{simplification}} the point cloud \(\X\), obtaining the point cloud \(\widetilde{\X}\), shown in Figure \ref{method}C.

The first persistent cohomology of \(\widetilde{\X}\) is then computed\footnote{See Appendix \ref{topological-background}.}, giving us a summary of the 1-dimensional holes in the data. We are interested in the longest living holes, as they represent stable features. In our analysis, we will focus on the longest living hole with lifespan \([a,b)\), and we want to understand how this hole is being reflected in the data. To do this, we will first pick a scale \(\epsilon \in [a,b)\) where this feature exists. We used \(\epsilon = a + 0.9(b-a)\) in our analysis\footnote{Our experience, which is shared with the authors of \cite{deSilva2011}, is that picking a scale near the end of the lifespan of the selected feature resulted in a better circular parametrization.}. Figure \ref{method}D shows the persistence diagram. The chosen feature \([a,b)\) is circled in red, and the scale \(\epsilon\) is marked with red lines. We will then construct a continuous map from the Rips complex of \(\widetilde{\X}\) at scale \(\epsilon\) to the circle \(S^1\) in such a way that the selected hole gets sent to the hole inside the circle. By restricting this map to the points of \(\widetilde{\X}\) we obtain a circular value, i.e. an \emph{angle} \(\theta(x)\) for each point \(x \in \widetilde{\X}\), as shown in Figure \ref{method}E. Intuitively speaking, the angle of a point indicates where the point lies relative to the selected hole. The process of constructing a circular map on \(\widetilde{\X}\) given a 1-dimensional hole is called \emph{circular parametrization}\footnote{See Appendix \ref{cohomology} for a more detailed explanation.}, and uses the representation of the 1-dimensional cohomology group of a space as maps from the space to \(S^1\). We use an improvement on the original procedure by \cite{deSilva2011}, which gives better results in cases when the data are non-uniformly distributed around the hole. This improvement is described in Appendix \ref{cohomology} under \emph{Improved smoothing}. We then extend the function \(\theta\) to \(\X\) as shown in Figure \ref{method}F by giving a point \(\mathbf{x}_t \in \X\) the circular value of its closest point in \(\widetilde{\X}\). In the end we get a circular value \(\theta(\mathbf{x}_t)\), for each time point \(\delta t\), i.e. a time-dependent circular value as shown in Figure \ref{method}G.

To improve the decoding, we will use the following tactic. First we run the procedure described above using a large smoothing width of \(\sigma = 1000ms\) such that only features with slow dynamics remain, while arguably less interesting features such as local theta phase preferences are ignored. Having decoded the feature, we use an information theoretic measure to determine which cells in the population are selective for this feature \citep{skaggs93}. We then run the procedure again with only the spike trains of the selective neurons. This time using a smaller smoothing width of \(\sigma = 250ms\) to allow for a finer decoding.

\paragraph{Residual analysis}
Identification of a single feature from a neural recording is a first step. A single recording, however, could contain a mixture of cells responding to different features or even multiple features \citep{rigotti2013}. 
We therefore take an iterative approach, wherein features are identified using topological methods and then explained away by a statistical model to reveal any additional features, similar to \cite{spreemann15}. 
We use a generalized linear model (GLM) \citep{mccullagh1989generalized} to predict the neural activity given the decoded angle. We then subtract the predicted neural activity from the original spike trains, obtaining residual spike trains. We can then apply our decoding procedure to the residual spike trains to uncover remaining features in the data.

\section*{Results}

\begin{figure*}[!h]
\adjustbox{valign=b}{%
  \begin{minipage}[t]{11.4cm}
  \includegraphics[width=\textwidth]{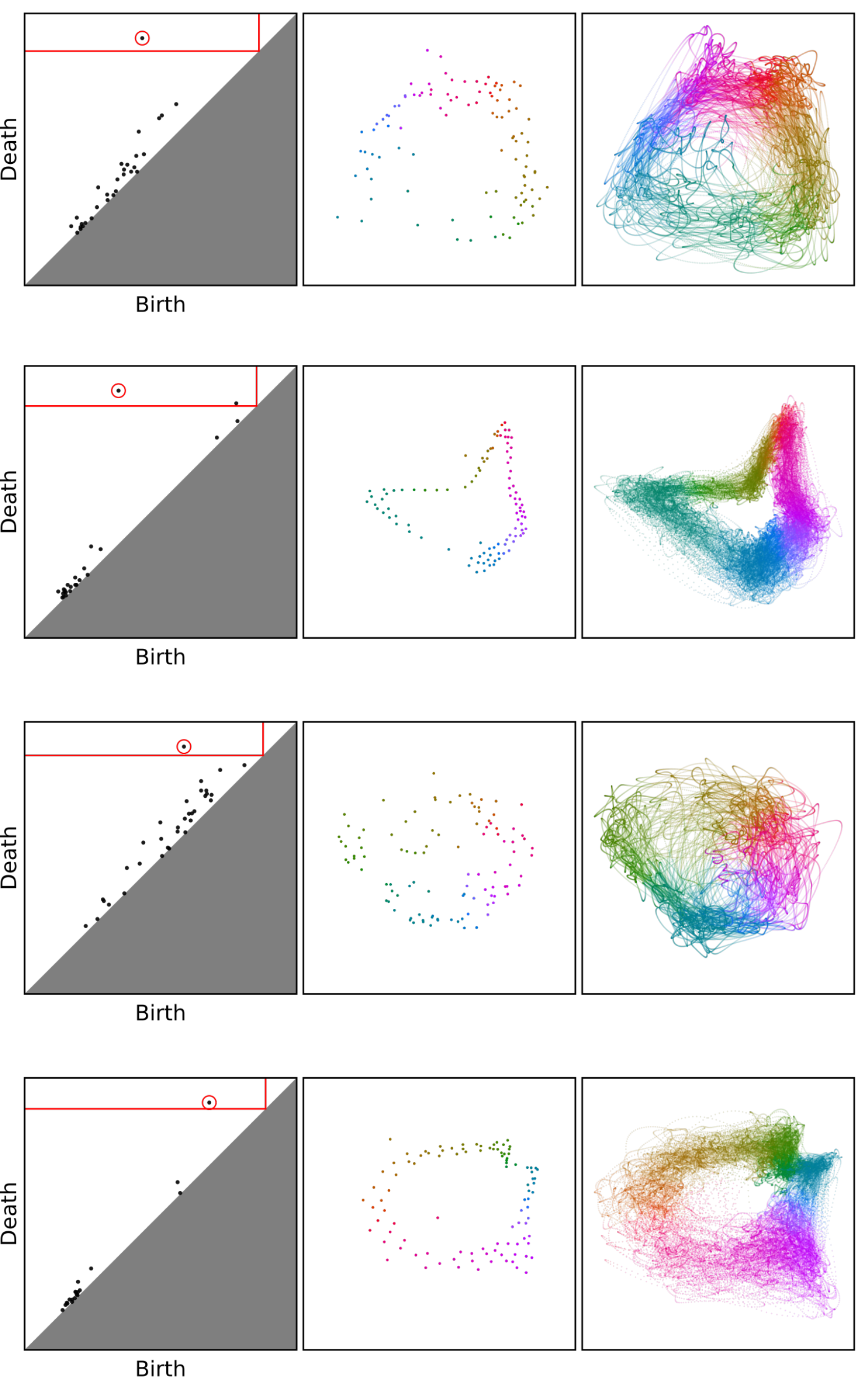}
\end{minipage}}%
\hspace{0.3cm}
  \adjustbox{valign=b}{%
    \begin{minipage}[t]{.34\linewidth}
      \caption{The decoding procedure applied to two recordings. From top to bottom: mouse28-140313 (using all neurons and using only identified selective neurons), mouse25-140130 (using all neurons and using only identified selective neurons). For each round we show the persistence diagram, the decoded circular value on the reduced point cloud and the circular value on the full point cloud. The point clouds in the second and third column are displayed as 2D projections.}
\label{results}
\end{minipage}
}
\end{figure*}
      
\paragraph{Rediscovering head direction cells}

\begin{figure*}[!ht]
\adjustbox{valign=b}{%
  \begin{minipage}[t]{11.4cm}
  \includegraphics[width=\textwidth]{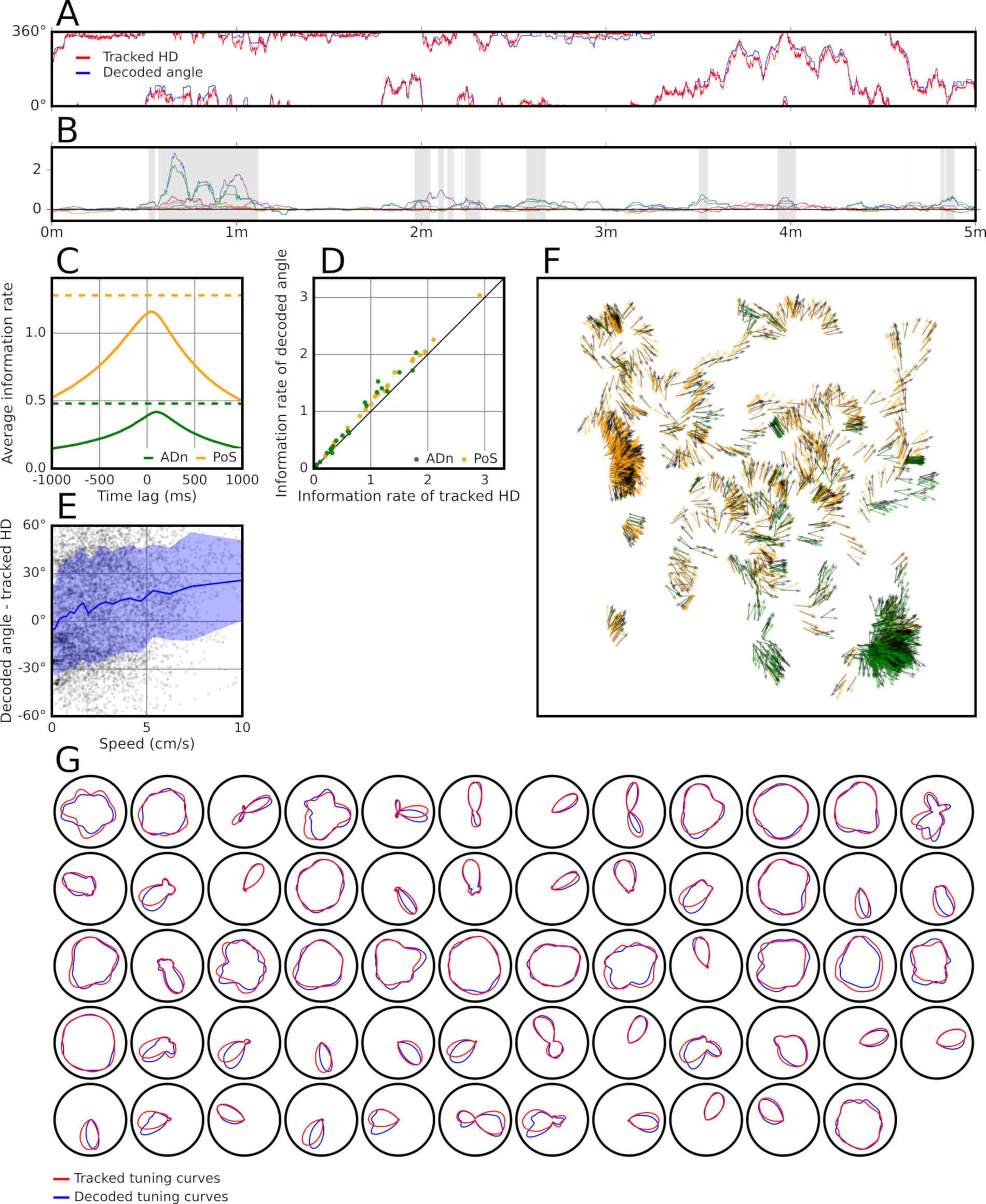}
\end{minipage}}%
\hspace{0.3cm}
  \adjustbox{valign=b}{%
    \begin{minipage}[t]{.34\linewidth}
      \caption{Result of the decoding procedure on mouse28-140313 (using only the identified selective neurons). A) Tracked HD and decoded angle for the first 5 minutes. B) Difference between log-likelihood of observed spike trains given decoded angle and tracked HD over individual neurons. The shaded time intervals represent the moments of drift. C) Average information rate over all ADn neurons (green lines) and PoS neurons (orange lines) of the tracked HD at different time lags (solid lines) and of the decoded angle (dashed lines) at 0 time lag. The average information rate of the tracked HD peaked at 95ms over ADn neurons and at 40ms over PoS neurons. D) Information rates of the tracked HD over all PoS neurons at 95ms time lag and over all ADn neurons at 40ms time lag, which is when their average information rates peaked, against information rates of the decoded angle at 0 time lag. E) Angular difference (decoded angle - tracked HD) for each time step during drift, plotted against the speed of the mouse. We also display the average angular difference (blue line) \(\pm 1\) SD (shaded region). F) For each time step during drift, we plot the tracked HD (black arrows) and the decoded angles, shown as a green (resp. orange) arrow when deviated to the clockwise (resp. counter-clockwise) of the tracked HD. The arrows are rooted at the position of the mouse at that time step. G) Tracked and decoded tuning curves of all neurons.}
\label{comparisons}
\end{minipage}
}
\end{figure*}

We applied our decoding procedure, as summarized in Figure \ref{method}, on spike train data from multielectrode array recordings of neurons in the antero-dorsal thalamic nucleus (ADn) and the post-subiculum (PoS) of seven freely moving mice \citep{peyrache15data,peyrache2015}. This revealed a prominent circular feature that was then decoded as shown in Figure \ref{results} and \ref{allresults}, resulting in a 1-dimensional circular time-dependent value. This decoded trajectory corresponded very well to the tracked head direction, as shown in Figure \ref{comparisons} and Video 1. 

In Figure \ref{comparisons}B, C and D we see that the decoded trajectories actually convey more information about the neural activity than the tracked head direction does. We were able to resolve moments of drift during the experiment, i.e. moments when the neural data are better explained by the decoded angle than by the tracked head direction. This was done by evaluating the time-varying log-likelihood ratio of two GLMs, one with the decoded angle and a second one with the tracked head direction. The difference of these two log-likelihoods are shown in Figure \ref{comparisons}B for each neuron. The GLM test found 90 moments of drift, lasting 2.01 seconds on average, that were dispersed throughout the experiment. Consistent with our findings, drift of the head direction representation has been previously reported in rodents \citep{taube90,yoder15,williams93,goodridge98}.

Figure \ref{comparisons}E shows that the discrepancy is mostly independent of the speed of the mouse, but with a slight clockwise deviation at slow speeds and a slight counter-clockwise deviation at higher speeds. In Figure \ref{comparisons}F we observe what appears to be a spatial dependence of the deviation, where the decoded angle is skewed counter-clockwise in some parts of the box and skewed clockwise in other parts. This suggests that the difference is not simply due to a random drift in the network representation \citep{zhang2112}, but rather that the internal representation is occationally distorted by the environment \citep{dabaghian2014,knierim425,peyrache17}. Another possible reason might be that the head direction is not precisely what is driving the neural activity but rather something similar such as gaze direction, the direction the body is facing or the direction that the animal is attending to. It could also be partially due to tracking error since the animal was tracked using a single camera at 30 frames per second and two LEDs on the animal's head.

\paragraph{Capturing speed cells}

\begin{figure*}[!ht]
\adjustbox{valign=b}{%
  \begin{minipage}[t]{11.4cm}
  \includegraphics[width=\textwidth]{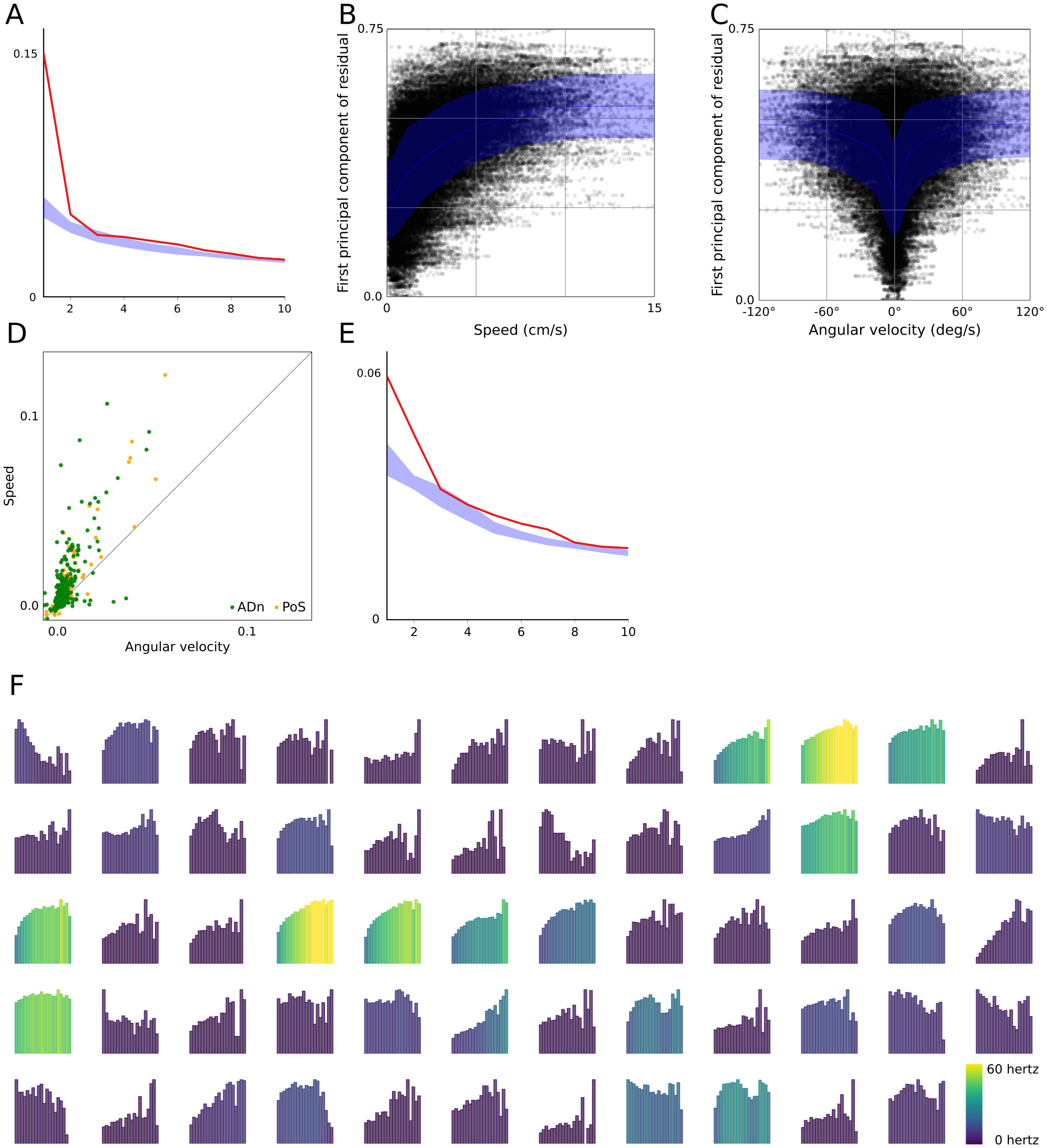}
\end{minipage}}%
\hspace{0.3cm}
  \adjustbox{valign=b}{%
    \begin{minipage}[t]{.34\linewidth}
      \caption{Residual analysis. A) The 10 largest eigenvalues (red line) of the residual point cloud after accounting for decoded HD, compared with a 96\% confidence interval of the 10 largest eigenvalues of each of 100 shuffled point clouds (see Appendix \ref{methods}). B and C) The first principal component of the residual point cloud against the speed (resp. angular velocity) of the mouse, with average values (blue line) \(\pm 1\) SD (shaded region) D) The pseudo \(R^2\)-score of a GLM including speed (x-axis) and a GLM including angular velocity (y-axis), averaged over a 5-fold cross-validation on all recorded neurons in the data set. E) The same as A but for the residual point cloud after accounting for decoded HD, speed and angular velocity. F) The speed tuning curves for all the neurons in mouse28-140313. On the x-axis of each tuning curve is the speed of the mouse, going from 0 cm/s to 15 cm/s. On the y-axis is the average firing rate of the neuron given the speed of the mouse. Note that the y-axes are scaled indepentently for each tuning curve.}
\label{residuals}
\end{minipage}
}
\end{figure*}

We then considered the residual activity remaining after accounting for the decoded angle in a GLM and applied the decoding procedure as before (shown in Figure \ref{residuals-supp}A). This time we did not find any additional cohomological features, but the eigenvalues of the covariance matrix of the residual point cloud suggest that there is a one-dimensional feature remaining in the data, as shown in Figure \ref{residuals}A. Two possible candidates are mouse speed and angular velocity, and Figure \ref{residuals}B and C show that both candidates are correlated with the neural activity. By fitting a GLM including speed and a GLM including angular velocity to the data we see in Figure \ref{residuals}D that most of the neurons are more selective for speed than for angular velocity. Finally, we included both decoded HD, speed and angular velocity in the GLM and created a new residual point cloud as before (shown in Figure \ref{residuals-supp}C). This time the eigenvalues (Figure \ref{residuals}E) are closer to random. We also tried to only include the residual spike trains of the neurons that had an initial mean firing rate higher than a certain threshold, varying the threshold between 0.05 and 20 spikes per second, but we were not able to uncover any structure in the remaining data (see supplementary figures \ref{firingrates} and \ref{extra-residual}).

\section*{Discussion}

As illustrated in our example, the main benefit of using persistent cohomology is that it allows us to understand the shape of the neural data. Statistical methods such as latent variable methods \citep{zhao2017, NIPS2014_5263, Frigola2014VariationalGP, Koyama2010ApproximateMF, Macke2011EmpiricalMO, Pfau2013RobustLO,Yu2008GaussianprocessFA} can give a good low-dimensional representation of neural activity, but they have not yet been developed to characterize the shapes underlying the data. It is often difficult, or even impossible, to identify the shape by looking at a 2D or 3D projection, as seen for instance in the first round on mouse25-140130 (Figure \ref{results}), and it would likely be even more difficult for more elaborate features. As a preprocessing step however, latent variable methods are far more flexible and powerful than PCA, allowing, for example, for known covariates and priors to be included simple and straightforward manner. Here we chose PCA for simplicity but see a bright future combining the benefits of cohomological feature extraction and the framework of latent variable models.

We emphasize that our method does not rely on knowing a priori what we are looking for, such as head direction. In our analysis, the tracked head direction is only used in the end to demonstrate that the method gives us the expected feature for this data. In practice, the discovery of a circular feature together with a circular parametrization would give the researcher useful knowledge of what kind of data is being encoded. For instance, if the circular value were instead moving in one direction on the circle with a certain frequency, this would suggest that something periodic is being encoded.

Topological data analysis has been previously applied to identify the shape of the underlying space in neural data \citep{singh2008,dabaghian2012,curto08}. This work, however, is the first to develop and apply topological methods for decoding the time-dependent variable. It is also a first example of using topological methods for systematic identification of relevant parameters, i.e. model selection \citep[see][for a demonstration on simulated data]{spreemann15}.

Some applications of topological methods to neural data \citep{spreemann15,curto08,giusti2015,dabaghian2012} have constructed spaces
based on correlations between neuron pairs.
This may work well when the tuning curves of the neurons are convex, but this assumption might be too strong. 
For instance, some of the neurons in Figure \ref{comparisons}G are selective for two different directions. 
This could be due to errors in preprocesing where the activity of two neurons are combined or might even be a property of the cells themselves. Our method does not require this assumption.

In our analysis we only focused on the longest living 1-dimensional hole, but in general this should depend on the situation. If more than one hole is considered significant, it would make sense to apply circular parametrization with respect to each feature. For instance, if the underlying space was a torus, persistent cohomology would detect two 1-dimensional features corresponding to the two orthogonal ways to traverse a torus, as well as one 2-dimensional feature. The 1-dimensional features would then give circular parametrizations that together determine the time-dependent position on the torus. Additional theoretical work, such as \cite{perea16}, should reveal the extent to which other features can be identified and decoded.

Ideally, when performing persistent cohomology, the real features are well separated from noise in the persistence diagram. For instance, in the diagrams in Figure \ref{results}, there is a prominent 1-dimensional hole. In other cases however, for instance some of the diagrams in Figure \ref{allresults}, it is not easy to separate the features from noise. However, even in these cases the circular parametrizations could still be meaningful, and our method of performing the decoding procedure again with only the neurons selective for the chosen feature might be able to remove some of the noise, making the interesting features more prominent.

Topological data analysis methods have developed into a powerful, intuitive set of tools, built with mathematical rigor, that can now be applied to large population neural recordings. Here we show one clear example of how these methods can recover the relevant features of the animal’s behavior, essentially performing unsupervised decoding and model selection. With the continued emergence of large population neural recordings, we expect to see topological methods playing an important role in exploring what the neural data are trying to tell us.

\newpage\phantom{blabla}
\newpage\phantom{blabla}

\appendix

\section{Topological background}
\label{topological-background}

The state of a neural population at a given time point can be represented as a high-dimensional vector \(x \in \R^n\). When collecting all the vectors corresponding to each time point we get what is called a \emph{point cloud} \(\X \subset \R^n\). Describing the shape of point clouds such as these, however, is non-trivial, but can be done using tools coming from the field of \emph{algebraic topology}.

Cohomology (see Chapter 3 of \cite{hatcher2002}) is one such tool, which is able to distinguish between shapes\footnote{Technically speaking the homotopy types.}, such as a circle, sphere or torus.
Intuitively, the 0-dimensional cohomology counts the number of connected components of the space, and for \(n\geq1\) the \(n\)-dimensional 
cohomology counts the number of \(n\)-dimensional holes in the space. 
For instance, a circle has one connected component and one 1-dimensional hole, a sphere has one connected component and one 
2-dimensional hole, and a torus has one connected component, two 1-dimensional holes and one 2-dimensional hole.

If we try to naively compute the cohomology of a finite point cloud \(\X \subset \R^n\) we run into a problem however, since this is just a finite set 
of points with no interesting cohomology. 
Instead we consider the cohomology of the \(\epsilon\)-thickening\footnote{In practice, we replace the \(\epsilon\)-thickenings 
by approximations called \emph{Rips complexes}, see Appendix \ref{cohomology}.} of \(\X\), denoted by \(\X_\epsilon\), consisting of the points in \(\R^n\) that are closer than \(\epsilon\) to a point in \(\X\).

If we are able to find a suitable scale \(\epsilon\), we might recover the cohomology of the underlying shape. 
Often however, there is noise in the data giving rise to holes, and there is no way to separate the noise from the real features. 
The solution is to consider all scales rather than fixing just one scale. 
This is where \emph{persistent cohomology}\footnote{In topological data analysis, persistent \emph{homology} is often used instead, but we will need cohomology for the decoding step. We refer to \cite{ghrist08,edelsbrunner2010} for introductions to persistence, and \cite{ghrist2014,carlsson2009} for introductions to applied topology in general.} \citep{deSilva2011} enters the picture.
Persistent cohomology tracks the cohomology of the space \(\X_\epsilon\) as \(\epsilon\) grows from 0 to infinity. 
Such a sequence of growing spaces is called a \emph{filtration}. 
When \(\epsilon = 0\), the set \(\X_\epsilon\) is equal to \(\X\) and has no interesting cohomology. 
On the other hand, when \(\epsilon\) is large enough, the space \(\X_\epsilon\) becomes one big component with no holes. 
In between these two endpoints however, holes may appear and disappear.

\begin{figure}[H]
  \centering
  \includegraphics[width=0.45\textwidth]{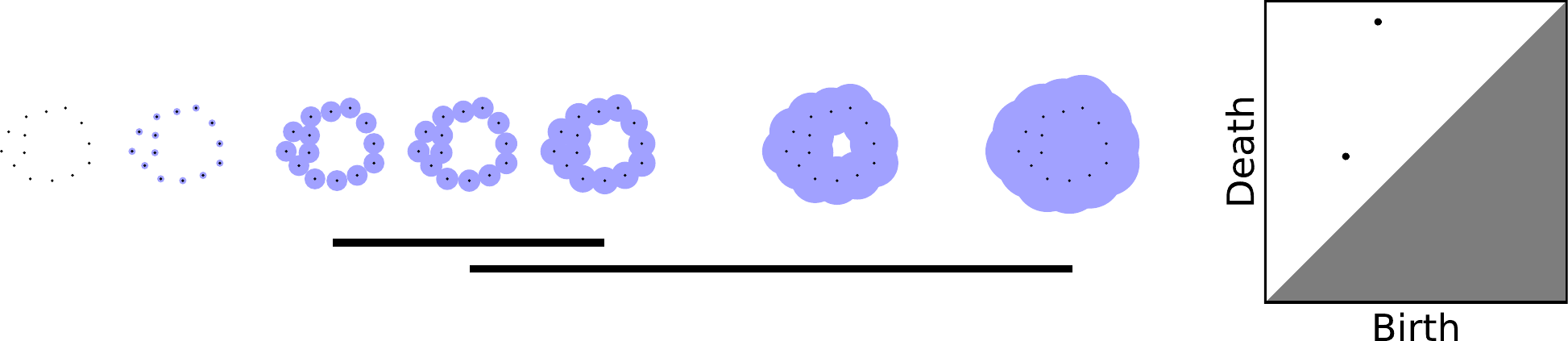}
  \caption{Left: A filtration together with its first persistent cohomology, drawn as a barcode. 
  Two 1-dimensional holes appear and disappear during the filtration, as reflected in the bars. Right: First persistence diagram of the filtration.}
  \label{filtration}
\end{figure}

The \(n\)'th persistent cohomology gives us the birth scale and death scale of all \(n\)-dimensional features. 
A feature with birth scale \(a\) and death scale \(b\) is denoted by \([a,b)\). 
The features can then be drawn as a \emph{barcode}, or alternatively, each feature \([a,b)\) can be drawn as the point \((a,b)\) in the plane, obtaining the \emph{\(n\)'th persistence diagram}, as shown in Figure \ref{filtration}. 
The features far away from the diagonal persist longer in the filtration, and are considered more robust, while features near the diagonal are considered as noise.

Once the persistence diagram has been computed, we want to relate the discovered features to the point cloud. 
This can be done using \emph{circular parametrization} (\cite{deSilva2011,johansson2015}, see also Appendix \ref{cohomology}), which turns any chosen 1-dimensional feature existing at a scale \(\epsilon\) into a circle-valued function \(f : \X \to S^1\), revealing the feature. See Figure \ref{method}.

\section{Details about the methods}

\paragraph{Identifying selective neurons}
We compute the information rate, as described in \cite{skaggs93}, for each neuron relative to the circular value using the formula \[I=\int_{S^1} \lambda(\theta)\,\log_2 \frac{\lambda(\theta)}{\lambda} p(\theta)\,d\theta,\] where \(\lambda(\theta)\) is the mean firing rate of the neuron when the circular value is \(\theta\), \(\lambda\) is the overall mean firing rate of the neuron, and \(p(\theta)\) is the fraction of time points where the circular value is \(\theta\). The integral is estimated by partitioning the circle in 20 bins of width \(18\) degrees, and assuming that the mean firing rate \(\lambda(\theta)\) is constant in each bin. We identify a neuron as being selective if its information rate is larger than 0.2. The method can then be run again without the spike trains of the non-selective neurons. The information rates of each neuron relative to the tracked HD, shown in Figure \ref{comparisons}C and D, were calculated in the same way.

\paragraph{Comparison with tracked HD}

For the comparisons with the tracked head direction in Figure \ref{comparisons}, we rotated the decoded circular value to minimize the mean squared deviation from the tracked HD. The tuning curves in Figure \ref{comparisons}G were made by taking the histogram of the angles at the spikes of each neuron, divide by the time spent in each angle, and smooth with a gaussian kernel of 10 degree standard deviation.

\paragraph{Calculating mouse speed and angular velocity}
The speed of the mouse was calculated by first smoothing the tracked position of the mouse separately in the x-axis and y-axis with a gaussian kernel of 0.1s, and then taking the central difference derivative of the smoothed position at each time step. Similarly, the angular velocity was calculated by first smoothing the tracked head direction of the mouse with a gaussian kernel of 0.1s, and then taking the central difference derivative at each time step.

\paragraph{GLM}
We fit a GLM to the activity of each neuron \(i \in \N\) as follows. First we binned the spike train of the neuron into bins of size \(25.6ms\), and the circle into angular bins of width \(36\) degrees. Let \(X_j(t)\) be the indicator function which is 1 if the decoded value is in the \(j\)'th angular bin at time step \(t\) and 0 otherwise, and let \(y_t\) be the observed spike count of the neuron in time bin \(t\). Assuming that the firing of the neuron is an inhomogeneous poisson process with instantaneous intensity at time step \(t\) given by \[H(t) = \sum^{9}_{j=0} \beta_j X_j(t) + h,\] where \(h\) and the \(\beta_j\) are parameters, the log-likelihood of observing the spike train of neuron \(i\) is given by \[L^{decoded}_i = \sum_t -H(t) + y_t \log(H(t)) - \log(y_t!).\] We used \cite{pyglmnet} to obtain the parameters \((h, \{\beta_j\})\) that maximizes the log-likelihood. The log-likelihood of the neuron having \(y_t\) spikes in bin \(t\) is now given by \[L^{decoded}_i(t) = -H(t) + y_t \log(H(t)) - \log(y_t!).\] We then repeated this process, replacing the function \(X_j(t)\) by the indicator function corresponding to the tracked HD, and obtained the corresponding values \(L^{tracked}_i(t)\). The difference \(L^{decoded}_i(t) - L^{tracked}_i(t)\) represents how much better the neuronal activity at time step \(t\) is explained by the decoded value compared to the tracked HD. This difference is shown for each neuron in Figure \ref{comparisons}B. The time intervals when the sum of these differences were above a certain threshold (we used \(thresh=1\)) is referred to as \emph{moments of drift}.

\paragraph{Residual analysis}
Given the above GLM, we made a residual spike train for each neuron \(i \in \N\) by taking the original spike train of the neuron and adding for each time step \(t\) a negative spike of magnitude \(H(t)\) at the center of bin \(t\). This is similar to the residual spike trains described in \cite{spreemann15}.

Figure \ref{residuals}A was made as follows. The red line shows the 10 largest eigenvalues of the covariance matrix of the residual point cloud after accounting for the decoded angle in the GLM. Each of the residual spike trains that gave rise to this point cloud were then shifted in time by \(n\) time steps, where \(n\) is a number sampled uniformly in the range from 0 to 100. This was done 100 times, resulting in 100 point clouds. For each of these point clouds, the 10 largest eigenvalues of their covariance matrix were chosen. The blue band in Figure \ref{residuals}B shows for each index \(i\) a \(96\%\) confidence interval of the \(i\)-th largest eigenvalue, sampled from this empirical distribution. 

\begin{figure*}[!hb]
\adjustbox{valign=b}{%
  \begin{minipage}[t]{11.4cm}
  \includegraphics[width=\textwidth]{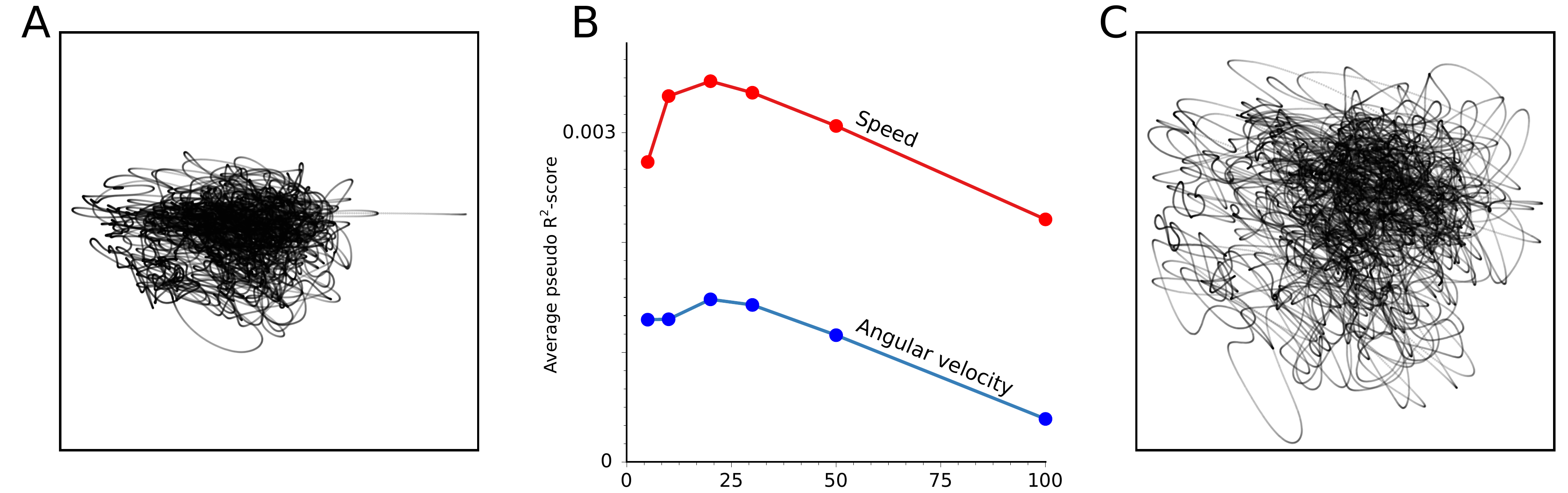}
\end{minipage}}%
\hspace{0.3cm}
  \adjustbox{valign=b}{%
    \begin{minipage}[t]{.34\linewidth}
      \caption{Supplementary figures to residual analysis in Figure \ref{residuals}. A) Residual point cloud after accounting for the decoded angle. B) The pseudo \(R^2\)-score of a GLM including speed (red line) and a GLM including angular velocity (blue line) for different number of bins, averaged over a 5-fold cross-validation on all recorded neurons in the data set. C) Residual point cloud after accounting for decoded angle, speed and angular velocity.}
\label{residuals-supp}
\end{minipage}
}
\end{figure*}

Figure \ref{residuals}D and \ref{residuals-supp}B were made as follows. We binned the mouse speed in \(n-1\) equally sized bins in the range \([0,15)\) and one bin \([15,\infty)\), and the angular velocity in \(n-2\) equally sized bins in the range \([-\frac{\pi}{2},\frac{\pi}{2})\) and the two bins \((-\infty, \frac{\pi}{2})\) and \([\frac{\pi}{2},\infty)\). We then defined two GLMs as before letting the functions \(X_j(t)\) be the indicator functions on the speed bins (resp. the angular velocity bins). These two models were cross-validated using 5 folds on each recorded neuron of every mouse in the data set, for different choises of the number of bins \(n\). The average pseudo \(R^2\)-score over all folds and over all neurons is shown in Figure \ref{residuals-supp}B for the two models and for different values of \(n\). The average pseudo \(R^2\)-score over all folds for each indivual neuron is shown in Figure \ref{residuals}D for the two models, using 20 bins in each model, which is where the average pseudo \(R^2\)-scores peaked.

Figure \ref{residuals}E was made the same way as Figure \ref{residuals}A, but for the residual point cloud after accounting for decoded angle, speed and angular velocity in the GLM.

\section{Point cloud simplification}
\label{simplification}

The procedure of simplifying the point cloud \(\X\) to obtain the point cloud \(\widetilde{\X}\) consists of two steps; \emph{Radial distance} and \emph{Topological denoising}.

\paragraph{Radial distance}
To make the computations faster, we simplify \(\X\) using the following method, called \emph{Radial distance} \citep{koning2011}. Start with the first point in \(\X\) and mark it as a key point. All consecutive points that have a distance less than a predetermined distance \(\epsilon\) to the key point are removed. The first point that have a distance greater than \(\epsilon\) to the key point is marked as the new key point. The process repeates itself from this new key point, and continues until it reaches the end of the point cloud. This procedure results in a smaller point cloud \(\X'\) that is close to the original point cloud\footnote{Since each point in \(\X'\) has a distance less than \(\epsilon\) to a point in \(X\), the Hausdorff distance between the two point clouds is less than \(\epsilon\).}. The parameter \(\epsilon\) is typically determined proportionally to the spread of the point cloud. In our analysis, we used \(\epsilon = 0.02\).

\paragraph{Topological denoising}

Since the point cloud is noisy, we need to reduce the amount of noise before we can look for topological features. We use a method for topological denoising that was introduced in \cite{carlsson09}. Given a subset \(S\) of \(\X'\), we define the function \[F(S,x) = \frac{1}{|\X'|} \sum_{p \in \X'} e^{\frac{-||x-p||^2}{2\sigma^2}} - \frac{\omega}{|S|} \sum_{p \in S} e^{\frac{-||x-p||^2}{2\sigma^2}}.\] The parameter \(\sigma\) is an estimate on the standard deviation of the noise, and the parameter \(w\) determines how much the points in \(S\) should repell each other. We maximize the function \(F\) by iteratively moving each point in \(S\) in the direction of the gradient of \(F\): Starting with a sample \(S_0 \subset \X'\), we let \[S_{n+1} = \left\{p + c \frac{\nabla F(S_n,p)}{M}\right\},\] for all \(n\), where \[M = \max_{p \in S_0} {||\nabla F(S_0,p)||}\]
and \(c\) is a parameter determing the maximum distance the points in \(S_n\) can move. In our case, we constructed \(S_0\) by taking every \(k\)'th point of \(\X'\), where \(k\) is the largest number such that \(S_0\) had 100 points. We used the parameters \footnote{The parameters used in our analysis were chosen experimentally, but automatical methods could be used instead.} \(\sigma = 0.1s, \omega = 0.1\) and \(c = 0.05\), where \(s\) is the standard deviation of \(\X'\). We did 60 iterations resulting in the point cloud \(\widetilde{\X}=S_{60}\).

\section{Circular parametrization}
\label{cohomology}

The material in this appendix is mostly a reformulation of parts of \cite{deSilva2011} with exception of the section \emph{Improved smoothing}, which is our own contribution.

\subsection*{Simplicial complexes}

\begin{figure}[H]
  \centering
  \includegraphics[width=0.45\textwidth]{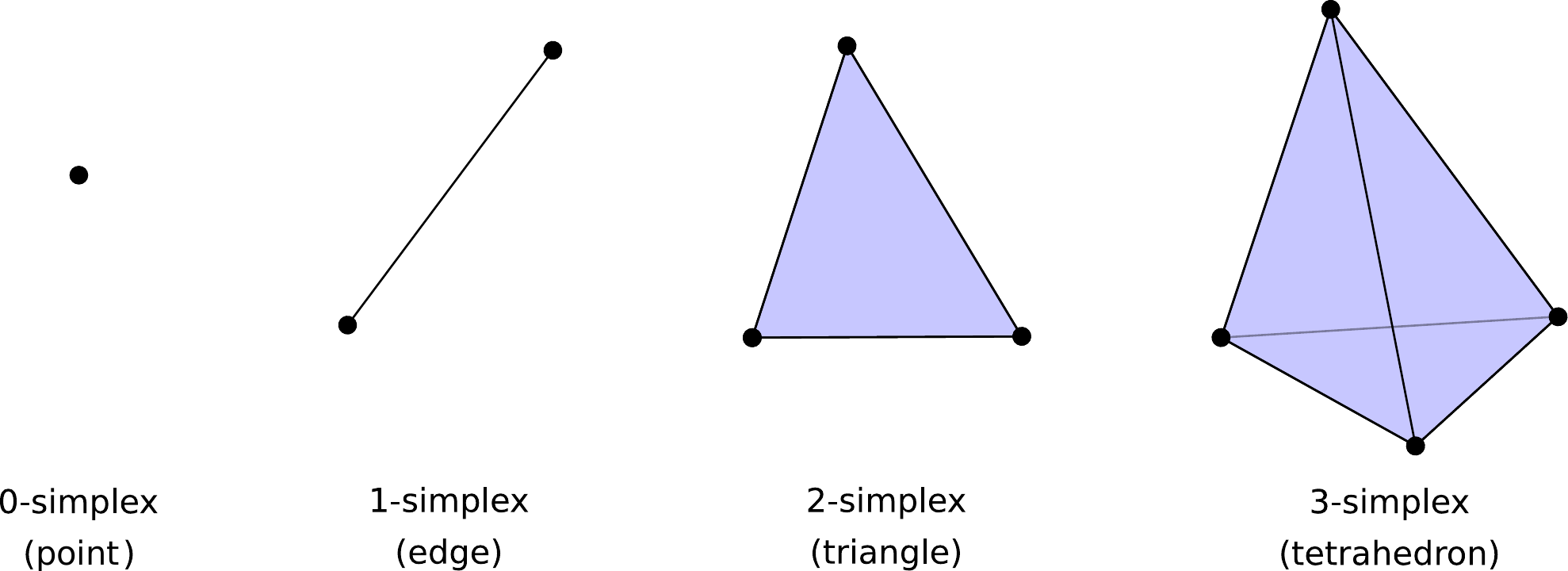}
  \caption{Simplices}
  \label{simplices}
\end{figure}

In topological data analysis, shapes are modeled by \emph{simplicial complexes}. A simplicial complex can be thought of as a space that is constructed by gluing together basic building blocks called simplices, as shown in Figure \ref{simplices}.

\begin{defn}
  A \emph{simplicial complex} is a pair \((X,S)\), where \(X\) is a finite set called the \emph{underlying set} and \(S\) is a family of subsets of \(X\), called \emph{simplices}, such that
  \begin{enumerate}
    \item For every simplex \(\sigma \in S\), all of its subsets \(\sigma' \subseteq \sigma\) are also simplices.
    \item For every element \(x \in X\), the one element set \(\{x\}\) is a simplex.
  \end{enumerate}
\end{defn}

An \emph{\(n\)-simplex} is a simplex consisting of \(n+1\) elements. The 0-simplices, 1-simplices and 2-simplices are called respectively points, edges and triangles. The simplicial complexes used in our method are called \emph{Rips complexes}, which are defined as follows.

%
%
%
%
%
%

\begin{defn}
  Let \(\X \subset \R^n\) be a point cloud. The \emph{Rips complex} of \(\X\) at scale \(\epsilon\), denoted \(R_\epsilon (\X)\) is the simplicial complex defined as follows:
  \begin{itemize}
    \item The underlying set is \(\X\).
    \item A subset \(\sigma \subset \X\) is a simplex iff \(||x-y|| \leq \epsilon\) for all \(x,y \in \sigma\).
  \end{itemize}
\end{defn}

%
%

\subsection*{Cohomology}

Let \(X\) be a simplicial complex, and let \(X_0\), \(X_1\) and \(X_2\) be respectively the points, edges, and triangles of \(X\). 

We will assume a total ordering on the points in \(X\). If \(\{a,b\}\) is an edge with \(a < b\), we will write it as \(ab\). Similarly, if \(\{a,b,c\}\) is a triangle with \(a < b < c\), we will write it as \(abc\).

Given a commutative ring \(\A\), for instance \(\Z\), \(\R\) or \(\F_p\), we define 0-chains, 1-chains and 2-chains with coefficients in \(\A\) as follows:

\[C^0 = C^0(X;\A) := \{f : X_0 \to \A\}\]
\[C^1 = C^1(X;\A) := \{\alpha : X_1 \to \A\}\]
\[C^2 = C^2(X;\A) := \{A : X_2 \to \A\}\]

These are modules over \(\A\). We define the coboundary maps \(d_0 : C^0 \to C^1\) and \(d_1 : C^1 \to C^2\) as follows:

\[d_0(f)(ab) = f(b)-f(a)\]
\[d_1(\alpha)(abc) = \alpha(ab) + \alpha(bc) - \alpha(ac)\]

A 1-chain \(\alpha\) is called a \emph{cocycle} if \(d_1(\alpha) = 0\), and it is called a \emph{coboundary} if there exists a 0-chain \(f \in C^0\) such that \(\alpha = d_0(f)\). It is easy to show that all coboundaries are cocycles. We define the first cohomology of \(X\) with coefficients in \(\A\) to be the module \[H^1(X;\A) = \frac{\kernel(d_1)}{\image(d_1)}.\] Two 1-chains \(\alpha\) and \(\beta\) are said to be \emph{cohomologous}, or belonging to the same \emph{cohomology class} if \(\alpha - \beta\) is a coboundary.

\subsection*{Circular parametrization}

The idea behind circular parametrization comes from the following theorem.

\begin{thm}[A special case of Theorem 4.57 in \cite{hatcher2002}]
  There is an isomorphism \[H^1(X,\Z) \simeq [X,S^1]\] between the 1-dimensional cohomology classes with integer coefficients of a space \(X\) and the set of homotopy classes of maps from \(X\) to the circle \(S^1 = \R/\Z\).
\end{thm}

Given a representative cocycle \(\alpha \in C^1(X;\Z)\), the associated map \(\theta : X \to \R/\Z\) is given by sending all the points in \(X\) to 0, sending each edge \(ab\) around the circle with winding number \(\alpha(ab)\), and extending linearly to the rest of \(X\). This extention is well-defined since \(d_1(\alpha) = 0\).

The circular maps obtained this way are not very smooth, since all points in \(X\) are sent to the same point on the circle. In order to allow for smoother maps, we consider cohomology with real coefficients. Consider \(\alpha\) as a real cocycle, and suppose we have another cocycle \(\beta \in C^1(X;\R)\) cohomologous to \(\alpha\). Since it is cohomologous, it can be written as \(\beta = \alpha + d_0(f)\) for some \(f \in C^0(X;\R)\). We define the map \(\theta : X \to \R/\Z\) by sending a point \(a\) to \(\theta(a) = f(a) \Mod{\Z}\), and an edge \(ab\) to the interval of length \(\beta(ab)\) starting at \(\theta(a)\) and ending at \(\theta(b)\). This map is extended linearly as before to the higher simplices of \(X\).

Given a cocycle \(\alpha \in C^0(X;\Z)\), we now want to find a real 0-chain \(f \in C^0(X;\R)\) such that the cocycle \(\beta = d_0(f)+\alpha\) is smooth, meaning that the edge lengths \(||\beta(ab)||\) are small. Define \[||\beta||^2 = \sum_{ab \in X_1} ||\beta(ab)||^2.\] We want to minimize this value. The desired 0-chain can be expressed as
\begin{equation}
  \label{original}
  f = \argmin_{\bar{f}} ||d_0(\bar{f}) + \alpha||.
\end{equation}
This is a least squares problem and can be solved using for instance \cite{lsqr}.

\subsection*{Improved smoothing}
\begin{figure}[H]
  \captionsetup[subfigure]{justification=centering}
  \begin{minipage}[t]{0.25\textwidth}
    \begin{subfigure}{\linewidth}
  \centering
    \includegraphics[width=0.9\textwidth]{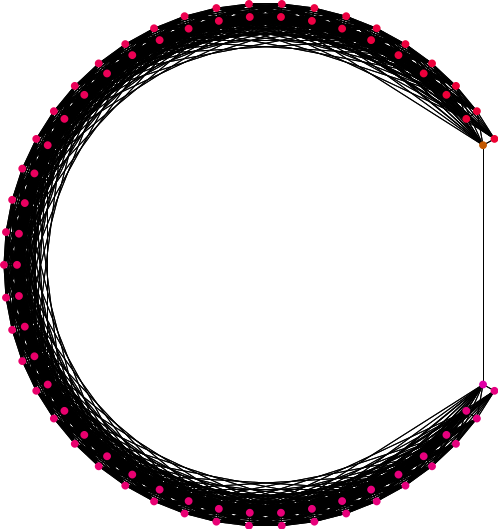}
    \caption{Original smoothing}
  \end{subfigure}%
    \begin{subfigure}{\linewidth}
  \centering
    \includegraphics[width=0.9\textwidth]{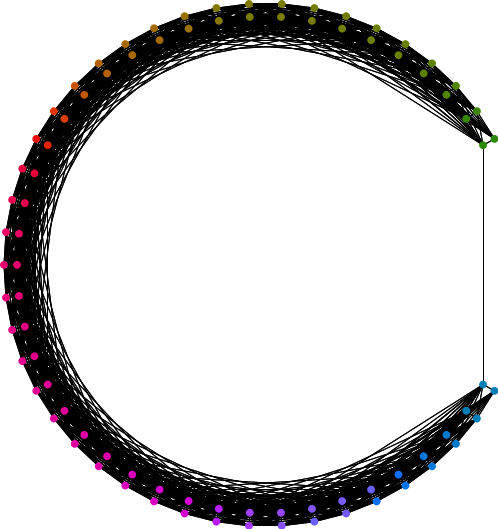}
    \caption{Improved smoothing}
  \end{subfigure}
  \end{minipage}
  \caption{Comparison between original and improved smoothing for the constructed example.}
  \label{constructedData}
\end{figure}

When we tried to apply circular parametrization to the data sets, we quickly discovered that it often produces unsatisfactory results. We demonstrate this using a constructed example, namely the Rips complex shown in Figure \ref{constructedData}a. The first cohomology of this complex with integer coefficient is \(\Z\), generated by the cocycle which has value 1 on the rightmost edge, and zero on all the other edges. The color of a point indicate its circular value after applying circular parametrization.


We see that the original smoothing sends the points to circular values that are close to each other. The reason for this behavior is that the edges in the simplicial complex are not evenly distributed around the circle; there is only one edge covering the rightmost part of the circle. Since this edge gives a relatively small contribution to the sum of squares of edge lengths, it will be streched around the circle so that the other edges can get smaller. This is problematic, because we want the geometry of the complex to be preserved.

We will describe a heuristical approach to improve the situation. First, we assume that the underlying set of the simplicial complex comes with a metric \(d : X \times X \to \R\). When the simplicial complex is a Rips complex of a point cloud \(\X \in \R^n\), we will take the metric to be the euclidean metric. Let \(\alpha \in C^1(X,\Z)\) be the cocycle that we want to use for circular parametrizaion. Now, instead of treating every edge equally, we will assign a positive weigth \(w : X_1 \to (0,\infty)\) to each edge in the simplicial complex. We define the weighting by the following procedure. First, we solve the original optimization problem (\ref{original}) to obtain a cocycle \(\beta = d_0(f) + \alpha\). We then construct a weighted directed graph \(G\) as follows: The vertices of \(G\) are the points \(a \in X^0\). For each edge \(ab \in X^1\) with \(\beta(ab) \geq 0\) there is an edge in \(G\) from \(a\) to \(b\) denoted \(ab\), and for each edge \(ab \in X^1\) with \(\beta(ab) < 0\) there is an edge in \(G\) from \(b\) to \(a\) denoted \(ba\). This gives us a bijection between edges in \(X\) and edges in \(G\). The weight of an edge \(ab\) is given by \(d(a,b)\). It can be shown that every edge \(ab\) in \(G\) has at least one directed cycle going through it. Now, for every edge in \(G\), take the shortest directed cycle in the graph going through it. This gives us a collection of cycles in \(G\). We now define \[w(ab) = \frac{l(ab)}{d(a,b)^2},\] where \(l(ab)\) is the number of cycles going through the corresponding edge in \(G\).

\begin{figure}[H]
  \begin{minipage}[t]{0.25\textwidth}
    \begin{subfigure}{\linewidth}
  \centering
    \label{doubleCircleOld}
    \includegraphics[width=0.9\textwidth]{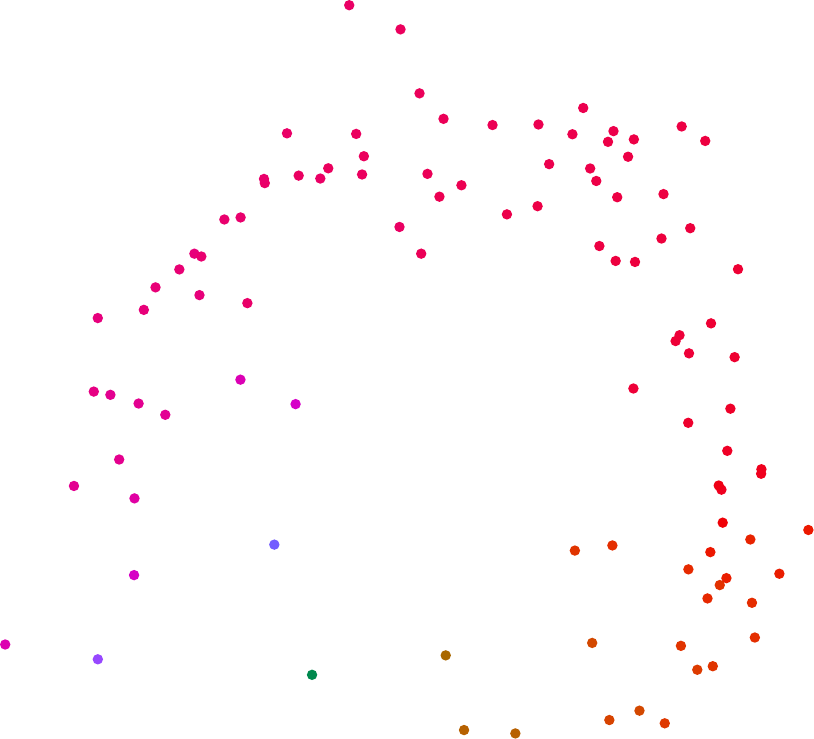}
    \caption{Original smoothing}
  \end{subfigure}%
    \begin{subfigure}{\linewidth}
  \centering
    \label{doubleCircleNew}
    \includegraphics[width=0.9\textwidth]{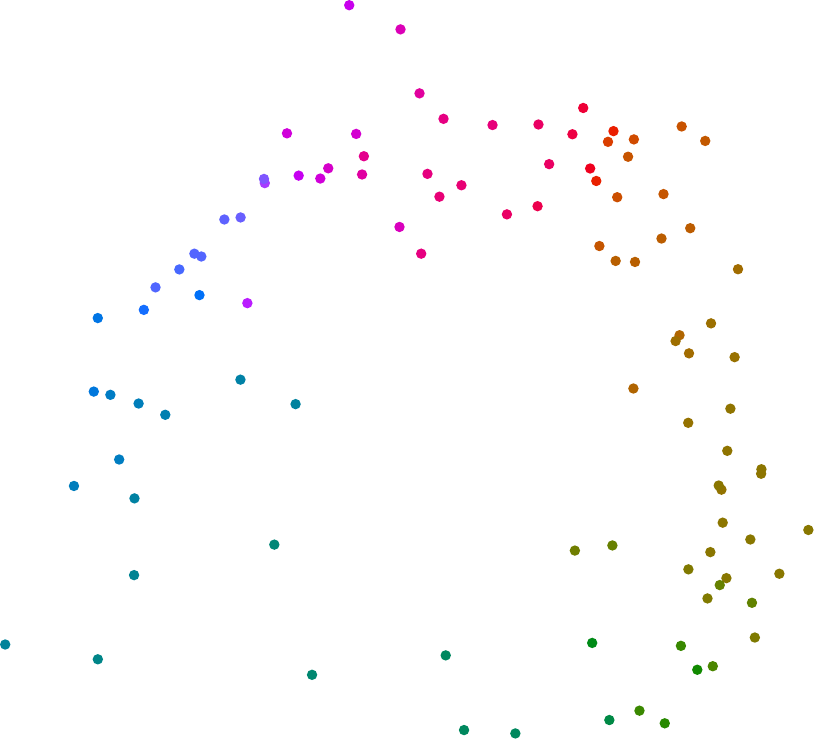}
    \caption{Improved smoothing}
  \end{subfigure}
  \end{minipage}
  \caption{Comparison between original and improved smoothing for real data.}
  \label{realData}
\end{figure}

Given this weighting, we then define \[||\beta||_w^2 = \sum_{ab \in X_1} w(ab)\,||\beta(ab)||^2.\] Our new optimization problem then becomes \[f = \argmin_{\bar{f}} ||d_0(\bar{f}) + \alpha||_w,\] which is again a least squares problem. After solving this system, we obtain a new circular parametrization. The result of applying this procedure to the Rips complex in our example is shown in Figure \ref{constructedData}b. Here the rightmost edge is given a much larger weighting than the other edges, since every directed cycle has to go through this edge. The result is that the circular values of the points are more evenly distributed around the circle. In Figure \ref{realData} we compare the original and improved smoothing on the circular parametrization done in the first round of mouse28-140313.

\section{Persistent cohomology}

By varying the parameter \(\epsilon\) of the Rips complex of a point cloud \(X\), we get a parametrized family of simplicial complexes \(R_\epsilon(\X)\), where \(\epsilon\) goes from 0 to infinity. For every pair of parameters \(0 \leq a \leq b\) we have a natural inclusion map \(i_a^b : R_a(\X) \to R_b(\X)\). We have the following two properties:
\begin{enumerate}
  \item The map \(i_a^a\) is the identity map on \(R_a(\X)\) for all \(a \in \R\).
  \item For all parameters \(a \leq b \leq c\) we have \(i_b^c \circ i_a^b = i_a^c\).
\end{enumerate}

Such a sequence of simplicial complexes together with maps satisfying the above two conditions is called a \emph{filtration}. Let \(p\) be a prime, and let \(\F_p\) be the field of order \(p\). When taking the cohomology of each space \(R_\epsilon(\X)\) with coefficients in \(\F_p\) we get a parametrized family of vector spaces \(H^n(R_\epsilon(\X))\) together with the induced linear maps \(\hat{i}_a^b = H^n(i_a^b) : H^n(R_b(\X)) \to H^n(R_a(\X))\) for every pair of parameters \(0 \leq a \leq b\). We have the following two properties:
\begin{enumerate}
  \item The map \(\hat{i}_a^a\) is the identity map on \(H^n(R_a(\X))\) for all \(a \in \R\).
  \item For all parameters \(a \leq b \leq c\) we have \(\hat{i}_a^b \circ \hat{i}_b^c = \hat{i}_a^c\).
\end{enumerate}

Such a sequence of vector spaces together with linear maps satisfying the above two conditions is called a \emph{persistence module}. The persistence module obtained by taking the cohomology of a filtration is called the \emph{persistent cohomology} of the filtration. Another kind of persistence modules are \emph{interval modules}.

\begin{defn}
  Let \(J\) be an interval on the real line. A \(J\)-interval module, written \(\I^J\) is the persistence module with vector spaces
  \[\I_a = \left\{
    \begin{array}{ll}
		\F  & \mbox{if } a \in J \\
    0 & \mbox{otherwise}
	  \end{array} \right.\]
  and linear maps
  \[f_a^b = \left\{
    \begin{array}{ll}
		I  & \mbox{if } a, b \in J \\
    0 & \mbox{otherwise}
	  \end{array} \right.\]
  where \(I\) denotes the identity map on \(\F\).
\end{defn}

\begin{thm}[A special case of Theorem 2.7 in \cite{chazal16}]
Given a finite point cloud \(\X\), the \(n\)th persistent cohomology of the Rips complex of \(\X\) can be decomposed into a sum of interval modules \[H^n(R(\X)) \simeq \I^{[a_1, b_1)} \oplus \I^{[a_2, b_2)} \oplus \dots \oplus \I^{[a_k, b_k)} = \I\]
\end{thm}

The \emph{persistent cohomology algorithm}, described in \cite{deSilva2011}, finds the intervals in the decomposition, along with a representative cocycle \(\alpha_{i,\epsilon} \in C^1(R_\epsilon(\X))\) for every interval \([a_i, b_i)\) and \(a_i \leq \epsilon < b_i\), such that 

\[\varphi_\epsilon ([\alpha_{i,\epsilon}]) = (0,\dots,0,\tikzmark{a}1,0,\dots,0).\]

\begin{tikzpicture}[remember picture,overlay]
\draw[<-] 
  ([shift={(2pt,-2pt)}]pic cs:a) |- ([shift={(-10pt,-10pt)}]pic cs:a) 
  node[anchor=east] {$\scriptstyle i\text{th position}$}; 
\end{tikzpicture}

By drawing each interval \([a, b)\) as a point \((a, b)\) in the plane we obtain the \emph{\(n\)th persistence diagram} of the filtration. We may now apply circular parametrization as follows. Pick an interval \([a_i, b_i)\) in the decomposition. Then choose a scale \(a_i \leq \epsilon < b_i\), and take the corresponding cocycle \(\alpha_{i,\epsilon}\). Now, this cocycle has coefficients in \(\F_p\), but circular parametrization requires a cocycle with integer coefficients. We need to \emph{lift} \(\alpha_{i,\epsilon}\) to an integer cocycle. We do this by first constructing an integer 1-chain by replacing each coefficient in \(\alpha_{i,\epsilon}\) with the integer in the same congruence class lying in the range \(\{-\frac{p-1}{2}, \dots \frac{p-1}{2}\}\). This almost always\footnote{If not, more work has to be done. See \cite{deSilva2011} for more details.} gives an integer cocycle \(\hat{\alpha}_{i,\epsilon} \in C^1(R_\epsilon(\X), \Z)\). We can then apply circular parametrization to this cocycle.

In our computations, we used Ripser \citep{ripser} to compute the first persistent cohomology with coefficients in \(\F_{47}\), giving persistence diagrams and the associated representative cocycles. The procedure described above to lift this cocycle always gave integer cocycles.

\section{Caption for Video 1 (attached)}

\textbf{Decoding on mouse28-140313.} We show the position of the two LED lights that were attached to the mouse (green and orange dots), the point in between the two lights (black dot), the tracked head direction (orange line) and the decoded angle (blue line) where we decoded using only the identified selective neurons.

\begin{figure*}[!hb]
  \includegraphics[width=\textwidth]{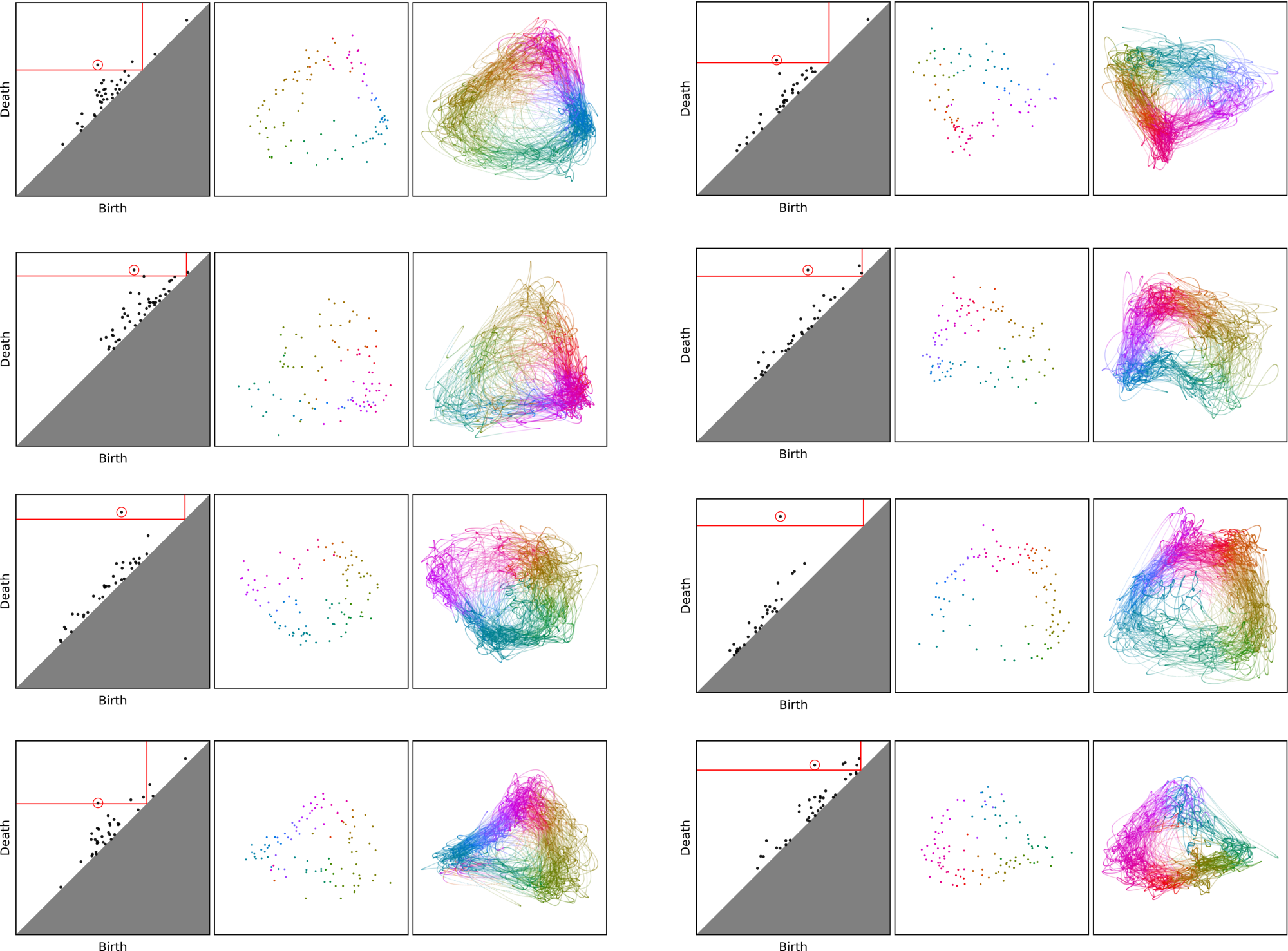}
\caption{The decoding procedure applied to different recordings, using all neurons.}
\label{allresults}
\end{figure*}

\begin{figure*}[h]
\hspace{2cm}
\adjustbox{valign=b}{%
  \begin{minipage}[t]{5cm}
    \includegraphics[width=\textwidth]{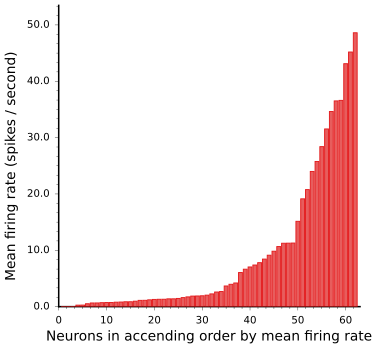}
\end{minipage}}%
\hspace{0.4cm}
  \adjustbox{valign=b}{%
    \begin{minipage}[t]{.34\linewidth}
      \caption{The mean firing rates of all 62 neurons of mouse28-140313.}
      \label{firingrates}
\end{minipage}
}
\end{figure*}

\begin{figure*}[!hb]
\includegraphics[width=0.95\textwidth]{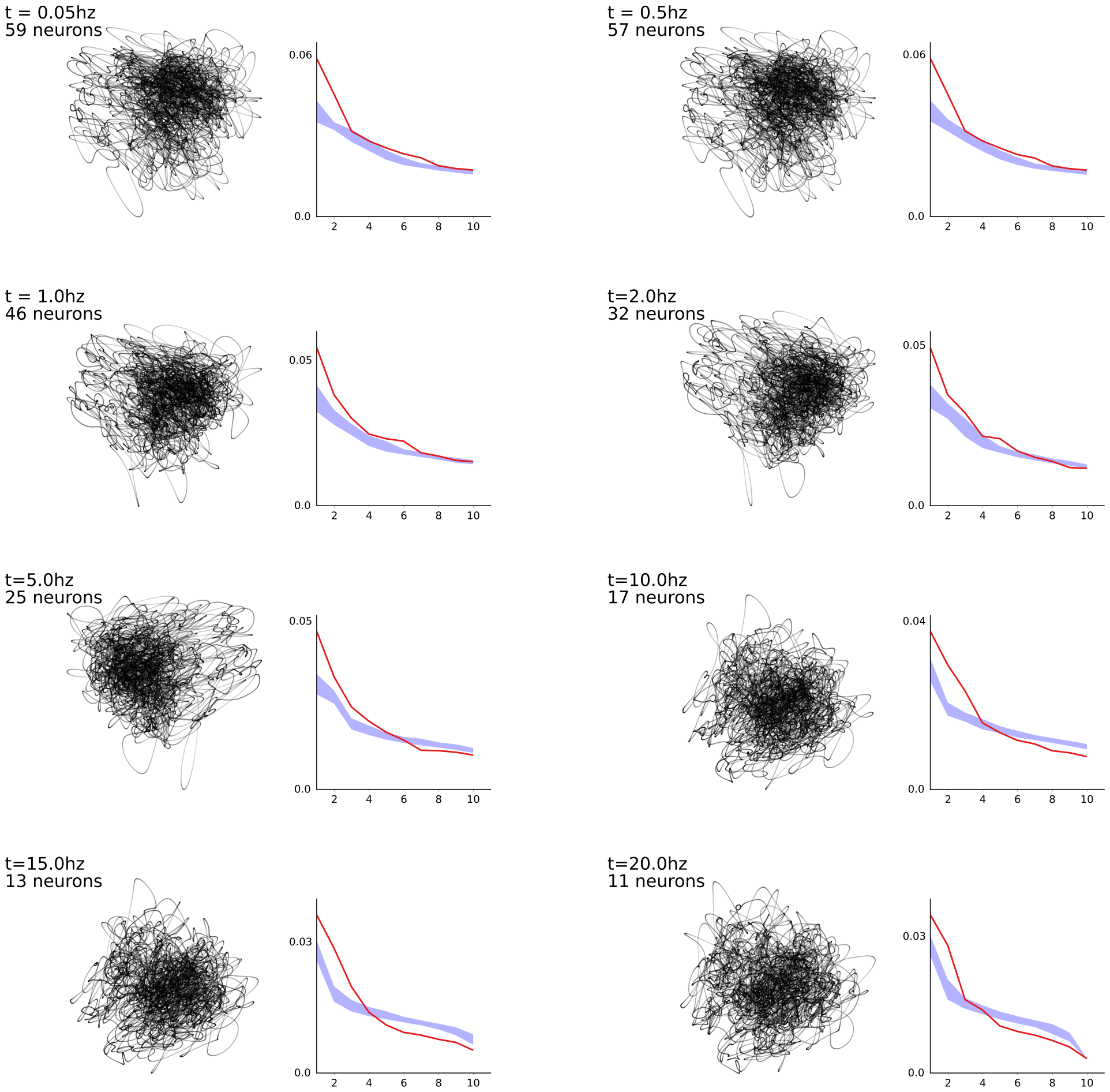}
\caption{We show for each threshold two figures. The first showing the resulting point cloud (in 2D) after removing the contribution of the decoded angle, speed and angular velocity, and the second showing the 10 largest eigenvalues (red line) of the residual point cloud compared with a \(96\%\) confidence interval of the 10 largest eigenvalues of each of 100 shuffled point clouds. We also show the threshold \(t\) and the number of neurons included.}
      \label{extra-residual}
\end{figure*}

\newpage\phantom{blabla}
\newpage\phantom{blabla}

\section{Supplementary figures}

\newpage\phantom{blabla}
\newpage\phantom{blabla}

\section*{Acknowledgement}
We would like to thank Adrien Peyrache and Gy{\"o}rgy Buzs{\'a}ki for sharing the data sets \citep{peyrache15data,peyrache2015} used in our analysis for free online.

\printbibliography

\end{document}